\documentclass[twocolumn,floatfix,superscriptaddress,showpacs,showkeys]{revtex4}
\usepackage{amsmath}
\usepackage{graphicx}
\usepackage{dcolumn}
\usepackage{bm}
\usepackage{booktabs}
\usepackage{amssymb}
\usepackage{multirow}
\usepackage{makecell}
\usepackage{textcomp}
\usepackage{subfigure}
\usepackage{phonetic}
\usepackage{extarrows}
\usepackage{color}
\usepackage{float}
\usepackage[colorlinks,citecolor=blue,linkcolor=blue]{hyperref}
\begin{document}


\title{Ground state cooling of magnomechanical resonator in $\mathcal{PT}$-symmetric cavity magnomechanical system at room temperature}

\author{Zhi-Xin Yang}
\affiliation{Department of Physics, College of Science, Yanbian University, Yanji, Jilin 133002, China}
\author{Liang Wang}
\affiliation{Department of Physics, College of Science, Yanbian University, Yanji, Jilin 133002, China}
\author{Yu-Mu Liu}
\affiliation{Center for Quantum Sciences and School of Physics, Northeast Normal University, Changchun 130024, China}
\author{Dong-Yang Wang}
\affiliation{School of Physics, Harbin Institute of Technology, Harbin, Heilongjiang 150001, China}
\author{Cheng-Hua Bai}
\affiliation{School of Physics, Harbin Institute of Technology, Harbin, Heilongjiang 150001, China}
\author{Shou Zhang}
\email{szhang@ybu.edu.cn}
\affiliation{Department of Physics, College of Science, Yanbian University, Yanji, Jilin 133002, China}
\author{Hong-Fu Wang}
\email{hfwang@ybu.edu.cn}
\affiliation{Department of Physics, College of Science, Yanbian University, Yanji, Jilin 133002, China}


\date{\today}

\begin{abstract}
We propose to realize the ground state cooling of magnomechanical resonator in a parity-time ($\mathcal{PT}$)-symmetric cavity magnomechanical system composed of a loss ferromagnetic sphere and a gain microwave cavity. In the scheme, the magnomechanical resonator can be cooled close to its ground state via the magnomechanical interaction, and it is found that the cooling effect in $\mathcal{PT}$-symmetric system is much higher than that in non-$\mathcal{PT}$-symmetric system. Resorting to the magnetic force noise spectrum, we investigate the final mean phonon number with experimentally feasible parameters and find surprisingly that the ground state cooling of magnomechanical resonator can be directly achieved at room temperature. Furthermore, we also illustrate that the ground state cooling can be flexibly controlled via the external magnetic field.
\end{abstract}

\maketitle


\section{\label{sec.1}Introduction}
In recent years, the developments on ferromagnetic systems which can achieve strong light-matter interactions have attracted widespread attention. The cavity magnomechanical systems, developed from cavity quantum electrodynamics (QED) systems, have been subjected to a rapid progress both in theory and experiment~\cite{Khitrova2006,PhysRevA.84.042339,PhysRevB.82.104413,PhysRevLett.116.223601,PhysRevLett.113.156401,PhysRevLett.117.133602,PhysRevLett.120.133602,PhysRevLett.121.087205,PhysRevLett.104.077202}. Among them, the researches on yttrium iron garnet (YIG) sphere, which is a kind of ferrimagnetic garnet material and is considered to be one of the most promising candidates for future quantum information processing, have been paid much attention due to the latest progress made by YIG sphere in the field of the development of magneto-optical technology. Owing to the unique physical characteristics, such as rich magnetic nonlinearity~\cite{PhysRevB.94.224410} and low damping rate~\cite{PhysRevLett.113.156401}, the Kittel mode (ferromagnetic resonance) of YIG sphere can realize the strong coupling~\cite{PhysRevLett.113.156401,PhysRevLett.111.127003} even ultrastrong coupling~\cite{PhysRevB.93.144420} between magnons and microwave cavity photons, which leads to the rapid development of the emerging field of cavity magnomechanical systems. On the other hand, the high spin density of YIG sphere leads to vacuum Rabi splitting and generates the quasiparticles, i.e., cavity magnon polarons\cite{PhysRevLett.113.156401,PhysRevLett.114.227201,PhysRevLett.113.083603,PhysRevLett.111.127003,PhysRev.73.155}. And in recent studies, because of its extraordinary robustness against to temperature, the coupling between magnons and cavity photons has been observed at both low temperature and room temperature~\cite{PhysRev.110.836,Zhang2015}. Many interesting phenomena have been investigated in cavity-magnon systems, such as the observation of bistability~\cite{PhysRevLett.120.057202}, magnon dark modes~\cite{Zhang2015}, higher-order exceptional point~\cite{PhysRevB.99.054404}, magnetically induced transparency~\cite{Wang:18}, magnetically manipulated slow light~\cite{Kong:19}, 
and other researches~\cite{PhysRevB.97.155129,PhysRevLett.114.227201,PhysRevApplied.2.054002,PhysRevA.99.043803}. Therefore, the cavity magnomechanical system provides us a new platform to study the strong coupling between magnons and microwave cavity photons and has achieved a major breakthrough in quantum information processing and exchange between qubits, for example, the couplings of magnon with superconducting qubits~\cite{Wallraff2004}, semiconducting qubits~\cite{PhysRevLett.106.247403}, phonons~\cite{PhysRevLett.5.100}, and so on. 
 
Currently, $\mathcal{PT}$-symmetric systems, which emerge as a peculiar platform to achieve distinctive optical behavior that is previously unattainable with general optical systems~\cite{Zhang2015}, have been deeply explored in the field of optics ~\cite{PhysRevA.96.023812,PhysRevA.94.043810,FrontiersOfPhysics.14.62601,PhysRevLett.106.093902,PhysRevLett.113.053604} and then been generalized to the magnon systems~\cite{AnnPhysBerlin.2000028,OptExpress.6.20248}. The $\mathcal{PT}$-symmetric systems show the important property when the $\mathcal{PT}$-phase transits from $\mathcal{PT}$-symmetric phase to broken $\mathcal{PT}$-symmetric phase~\cite{PhysRevX.6.021007,PhysRevLett.112.143903}. Furthermore, the optical non-reciprocity has been experimentally reported in $\mathcal{PT}$-symmetric whispering-gallery microcavities (WGM)~\cite{Peng2014}, and the measurement sensitivity of detecting weak mechanical motion can be uncommonly enhanced near the $\mathcal{PT}$-phase transition point (exceptional point)~\cite{PhysRevLett.117.110802}. 

To observe the quantum effect in a cavity magnomechanical system, it is a prerequisite to cool the magnomechanical resonator to its quantum ground state~\cite{PhysRev.110.836,PhysRevLett.99.093902,Teufel2011,PhysRevLett.99.093901,PhysRevA.91.033818,FrontPhys.6.237250,PhysRevA.94.033809,ding2019ground}. In this paper, inspired by the above, we investigate the ground state cooling of magnomechanical resonator with experimentally achievable parameters in a $\mathcal{PT}$-symmetrical magnomechanical system, which is consisted of a loss YIG sphere and a gain microwave cavity. Under the weak coupling regime ($G\ll\kappa_{a}$), we study the cooling effect of magnomechanical resonator in $\mathcal{PT}$-symmetric and non-$\mathcal{PT}$-symmetric systems based on the magnetic force noise spectrum, and find that in the $\mathcal{PT}$-symmetric system, the cooling effect is obviously enhanced because the heating process is greatly suppressed. Interestingly and surprisingly, we find that by calculating the final mean phonon number from the magnetic force noise spectrum, the magnomechanical resonator can be cooled directly to its ground state at room temperature without cryogenic precooling. Furthermore, we also explore the effect of bias magnetic field $H$ on the cooling of the magnomechanical resonator.
 
The rest of the paper is organized as follows. In Sec.~\ref{sec.2}, we illustrate the physical model and derive the effective Hamiltonian of the $\mathcal{PT}$-symmetric cavity magnomechanical system. In Sec.~\ref{sec.3}, by calculating the magnetic force noise spectrum and the final mean phonon number, we show how to realize the ground state cooling of the magnomrchanical resonator at room temperature with the experimentally feasible parameters, and reveal that the $\mathcal{PT}$-symmetry plays a key role in enhancing the cooling effect. Also, we study the effect of bias magnetic field on the cooling scheme. Finally, we summarize our work in Sec.~\ref{sec.4}.

\begin{figure}
	\centering
	\includegraphics[width=0.5\linewidth]{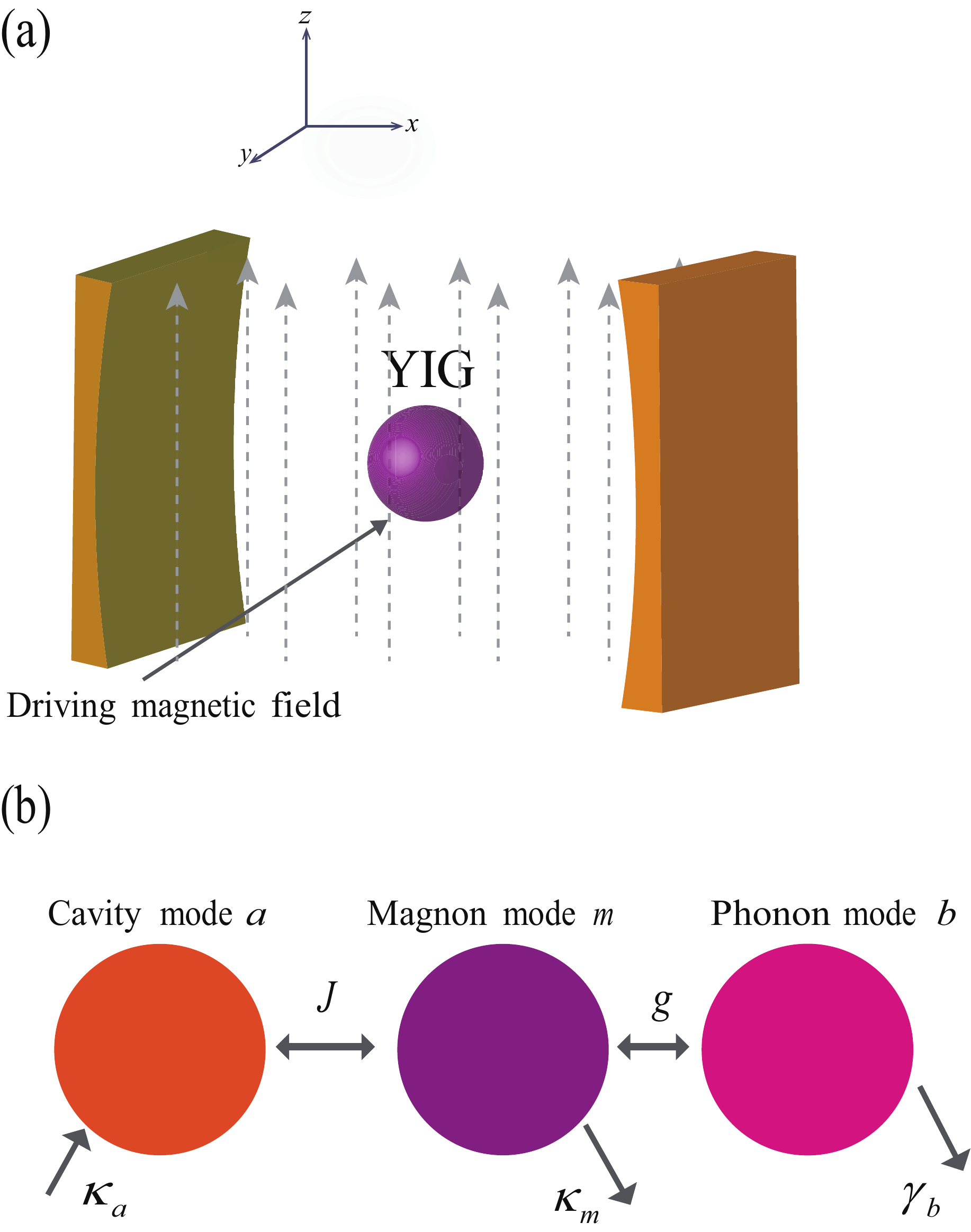}
	\caption{(a) Schematic diagram of the cavity magnomechanical system, which includes a gain microwave cavity and a loss YIG sphere. The magnetic field ($x$ direction) of the microwave cavity mode, the driving magnetic field ($y$ direction), and the bias magnetic field ($z$ direction) are mutually perpendicular at the site of the YIG sphere. (b) The system considered in the above is equivalent to a coupling-harmonic-resonator model. Here, $\kappa_{a}$ is the gain rate of microwave cavity mode $a$, $\kappa_{m}$ and $\gamma_{b}$ denote the decay rates of the magnon mode $m$ and phonon mode $b$, respectively, $J$ is the coupling strength between the magnon mode and the microwave cavity mode, and $g$ is the coupling strength between the magnon mode and the phonon mode.}
	\label{fig1}
\end{figure}

\section{\label{sec.2} Model And Hamiltonian}
We consider a hybrid cavity magnomechanical system, which consists of a loss YIG sphere (a 125-$\mu$m-radius sphere \cite{Zhange1501286}) and a gain microwave cavity, as shown in Fig.~\ref{fig1}(a). Due to its material and geometric characteristics, the loss YIG sphere placed in the gain cavity has both a magnetic mode and a mechanical mode. Usually, the loss YIG sphere is placed near the maximum microwave magnetic field of the gain microwave cavity mode, and a uniform bias magnetic field $H$, whose adjusting range is between 0 and 1 $\mathrm{T}$ \cite{PhysRevB.94.224410}, is introduced into the $z$-axis to achieve the magnon-photon coupling~\cite{PhysRevLett.120.057202}, which can be adjusted by changing the position of the YIG sphere in the microwave cavity. The magnetic field of the microwave cavity mode, the driving magnetic field, and the bias magnetic field are along the $x$, $y$, and $z$ directions, respectively. The change of the magnetization induced by the magnon excitation inside the YIG sphere causes the deformation of its geometry structure, which forms the vibrational modes (phonons) of the YIG sphere, and vice versa~\cite{PhysRev.110.836}. Thus, the YIG sphere can be regarded as a magnomechanical resonator. 

In Fig.~\ref{fig1}(b) we present the equivalent model, in which there are three modes in the system: mircrowave cavity photon mode, magnon mode, and phonon mode. The couplings between the magnon-photon and magnon-phonon modes can be achieved via the magnetic dipole interaction and magnetostrictive interaction, respectively. Since the size of the YIG sphere is much smaller than the wavelength of microwave cavity, the interaction between microwave cavity mode and phonon mode caused by radiation pressure thus can be neglected. The total Hamiltonian of the system is written as ($\hbar=1$)  
\begin{eqnarray}\label{e01}
H_{total}&=&
\omega_{a}a^{\dagger}a+\omega_{b}b^{\dagger}b+\omega_{m}m^{\dagger}m+J\left(ma^{\dagger}+m^{\dagger}a\right)
\cr\cr&&
+gm^{\dagger}m\left(b^{\dagger}+b\right)+i\Omega\left(m^{\dagger}e^{-i\omega_{L}t}-me^{i\omega_{L}t}\right),
\end{eqnarray}
where $a$ $(a^{\dagger})$, $b$ $(b^{\dagger})$, and $m$ $(m^{\dagger})$ are annihilation (creation) operators of the photon, phonon, and magnon modes, respectively; $\omega_{a},\omega_{b}$, and $\omega_{m}$ represent the resonance frequencies of the photon, phonon, and magnon modes, respectively; $J$ and $g$ are the coupling rates of the magnon-cavity interaction and magnon-phonon interaction. The magnetostrictive coupling $g$ is usually small in experiments~\cite{Zhange1501286}, but the  magnetostrictive interaction can be enhanced by driving the magnon mode by using a strong microwave field (the YIG sphere are directly driven by a microwave source, so the magnetostrictive coupling can be enhanced~\cite{PhysRevLett.121.203601}). The Rabi frequency $\Omega=\frac{\sqrt{5}}{4}\gamma_{g}\sqrt{M}B_{0}$ represents the external driving amplitude acting on the magnon mode~\cite{PhysRevLett.121.203601,PhysRev.58.1098,PhysRevLett.84.2726}, with $B_{0}$ and $\omega_{L}$ being the amplitude and the frequency of the driving magnetic field, respectively. The total number of spins $M=\rho V$, where $\rho=4.22\times10^{27}\mathrm{m}^{-3}$ is the spin density of the YIG sphere and $V$ is the volume of the sphere. In the rotating frame with respect to the driving frequency $\omega_{L}$, the Hamiltonian of the system is given by
\begin{eqnarray}\label{e02}
H^{'} &=&
-\Delta_{a}a^{\dagger}a+\omega_{b}b^{\dagger}b-\Delta_{m}m^{\dagger}m+J\left(m a^{\dagger}+m^{\dagger}a\right)
\cr\cr&& 
+gm^{\dagger}m\left( b^{\dagger}+b\right) +i\Omega\left( m^{\dagger}-m\right),
\end{eqnarray}
where $\Delta_{a} = \omega_{L}-\omega_{a}$ and $\Delta_{m} =\omega_{L}-\omega_{m}$ are the detunings of the microwave cavity mode and magnon mode, respectively. 

In order to study the dynamics of the cavity magnomechanical system, we apply the standard linearization process. By rewriting all bosonic operators as the sum of its steady state mean value and a small fluctuation term, i.e., $a=\alpha+\delta a$, $b=\beta+\delta b$, and $m=\zeta+\delta m$, where $\alpha$, $\beta$, and $\zeta$ denote the steady-state mean values of three modes, and $\delta a$, $\delta b$, and $\delta m$ are small fluctuation terms of three modes, respectively. The linearized Hamiltonian of the system is written as
\begin{eqnarray}\label{e03}
H_{L} &=&
-\Delta_{a}\delta a^{\dagger}\delta a+\omega_{b}\delta b^{\dagger}\delta b-\Delta^{'}_{m} \delta m^{\dagger}\delta m +J(\delta m\delta a^{\dagger}\cr\cr&& +\delta m^{\dagger}\delta a)
+\left( G\delta m^{\dagger}+G^{*}\delta m\right) \left(\delta b^{\dagger}+\delta b\right),
\end{eqnarray}
where $\Delta^{'}_{m}=\Delta_{m}-g(\beta^{*}+\beta)$ is the modified detuning of the magnon mode and $G=g\zeta$ is the enhanced magnetostrictive coupling strength via the coherent driving. Assume that $g(\beta^{*}+\beta)\ll\Delta_{m}$, we can approximately obtain $\Delta^{'}_{m}\approx\Delta_{m}$. By solving the steady state Langevin equations, $\beta = −ig \left| \zeta\right| ^2 /(i\omega_{b} +\gamma_{b}/2)$ can be obtained and $\zeta$ is given by
\begin{eqnarray}\label{e04}
\zeta&=&\frac{\Omega\left( -i\Delta_{a}+\kappa_{a}\right)}{J^2+\left( -i\Delta_{a}+\kappa_{a}\right) \left( -i\Delta_{m}+\kappa_{m}\right) },
\end{eqnarray}
where $\kappa_{a}$ is the gain rate of the microwane cavity mode, $\gamma_{b}$ and $\kappa_{m}$ are the decay rates of the phonon mode and magnon mode, respectively. Since $\zeta$ is affected by the driving field, we can enhance $G$ by adjusting the external driving field $\Omega$. 

The quantum Langevin equations of the system operators for the linearized Hamiltonian in Eq.~(\ref{e03}) are given by
\begin{eqnarray}\label{e05}
\delta\dot{a} &=& 
\left( i\Delta_{a}+\frac{\kappa_{a}}{2}\right)\delta a-iJ\delta m-\sqrt{\kappa_{a}}a_{\mathrm{in}},
\cr\cr
\delta\dot{b} &=& 
\left( -i\omega_{b}-\frac{\gamma}{2}\right)\delta b-i\left( G\delta m^{\dagger}+G^{*}\delta m\right) -\sqrt{\gamma_{b}}b_{\mathrm{in}},
\cr\cr
\delta\dot{m} &=& 
\left( i\Delta_{m}-\frac{\kappa_{m}}{2}\right)\delta m-iJ\delta a-iG\left(\delta b^{\dagger}+\delta b\right)-\sqrt{\kappa_{m}}m_{\mathrm{in}},
\end{eqnarray}
where $b_{\mathrm{in}}$ and $m_{\mathrm{in}}$ are the corresponding noise operators with zero mean value, with the nonzero correlation functions,
\begin{eqnarray}\label{e06}
\left\langle b^{\dagger}_{\mathrm{in}}(t)b_{\mathrm{in}}(t^{'})\right\rangle &=& n_{\mathrm{th}}\delta( t-t^{'}),
\cr\cr
 \left\langle b_{\mathrm{in}}(t) b^{\dagger}_{\mathrm{in}}(t^{'})\right\rangle &=&(n_{\mathrm{th}}+1)\delta(t-t^{'}),
 \cr\cr
 \left\langle m^{\dagger}_{\mathrm{in}}(t)m_{\mathrm{in}}(t^{'})\right\rangle &=&0, 
 \cr\cr
 \left\langle m_{\mathrm{in}}(t) m^{\dagger}_{\mathrm{in}}(t^{'})\right\rangle &=&\delta(t-t^{'}).
\end{eqnarray}
Here $n_{\mathrm{th}}=[\mathrm{exp}(\hbar\omega_{b}/k_{\mathrm{B}}T)-1]^{-1}$ is thermal phonon number of the mechanical resonator, where $k_{\mathrm{B}}$ is the Boltzmann constant and $T$ is the environmental temperature. For the gain microwave cavity mode, the intrinsic quantum noise is described via the noise operators $a_{\mathrm{in}}$ and $a^\dagger_{\mathrm{in}}$ , which satisfy~\cite{PhysRevA.91.053832,PhysRevA.91.033830,PhysRevA.85.031802,PhysRevA.94.031802,Kepesidis_2016}.
\begin{eqnarray}\label{e07}
\left\langle a^{\dagger}_{\mathrm{in}}(t) a_{\mathrm{in}}( t^{'}) \right\rangle &=&\delta(t-t^{'}) , 
\cr\cr
\left\langle a_{\mathrm{in}}(t) a^{\dagger}_{\mathrm{in}}( t^{'}) \right\rangle &=&0.
\end{eqnarray} 
The effects of temperature on the microwave cavity mode and magnon mode can be ignored when the frequencies are very high. In other word, if the environment of the cavity is assumed to be in the vacuum state, the correlation functions of the noise operators for the gain cavity are independent of the environmental temperature $T$ .

Due to the presence of gain, the mode splitting and linewidths of the supermodes will be changed. Before proceeding, it is important to briefly explain the main mechanism behind $\mathcal{PT}$-symmetry in the present system. In the following, we discuss and analyze the effect of phase transition point of $\mathcal{PT}$-symmetry system on the ground state cooling of the magnomechanical resonator, which distinguishes the dynamical phenomena of unbroken and broken $\mathcal{PT}$-symmetric regions.

If we consider only the microwave cavity mode and magnon mode, the effective non-Hermitian Hamiltonian is written as
\begin{eqnarray}\label{e08}
H_{eff}&=&\omega_{a}\delta a^{\dagger}\delta a+ \omega_{m}\delta m^{\dagger}\delta m+\frac{i\kappa_{a}}{2}\delta a^{\dagger}\delta a\cr\cr&&-\frac{i\kappa_{m}}{2}\delta m^{\dagger}\delta m
+J(\delta a^{\dagger}\delta m+\delta m^{\dagger}\delta a),
\end{eqnarray}
which includes the gain rate of the cavity mode and the decay rate of the magnon mode. Assume that $\omega_{a}=\omega_{m}=\omega_{0}$, the effective Hamiltonian can be rewriten as
\begin{eqnarray}\label{e09}
H_{eff}&=&\xi_{+}B^{\dagger}_{1}B_{1}+\xi_{-}B^{\dagger}_{2}B_{2},
\end{eqnarray}
where $B_{1}=(\delta m+\delta a)/\sqrt{2}$ and $B_{2}=(\delta m-\delta a)/\sqrt{2}$ are two supermode operators, with the corresponding eigenfrequencies 
\begin{eqnarray}\label{e010}
\xi_{\pm}&=& \omega_{0}-\frac{i\chi}{2}\pm\sqrt{J^{2}-\Gamma_{eff}^{2}},
\end{eqnarray}
here $\Gamma_{eff}=(\kappa_{a}+\kappa_{m})/4$ and $\chi=(\kappa_{m}-\kappa_{a})/2$, and the supermodes $B_{1}$ and $B_{2}$ correspond to the frequencies $\xi_{+}$ and $\xi_{-}$, respectively. In our system, if the effective Hamiltonian of cavity-magnon system is unchanged under the time and parity reversal transformations, the system thus can be called $\mathcal{PT}$-symmetric system. To obtain the $\mathcal{PT}$ symmetric system, two necessary conditions should be satisfied: ($\textrm{i}$) the magnon mode and the microwave cavity mode have the same frequency, i.e., $\omega_{a}=\omega_{m}=\omega_{0}$; ($\textrm{ii}$) the gain rate of the microwave cavity mode and the decay rate of the magnon mode are balanced, i.e. $\kappa_{a}=\kappa_{m}$.
 \begin{figure}
	\centering
	\subfigure{\includegraphics[width=0.48\linewidth]{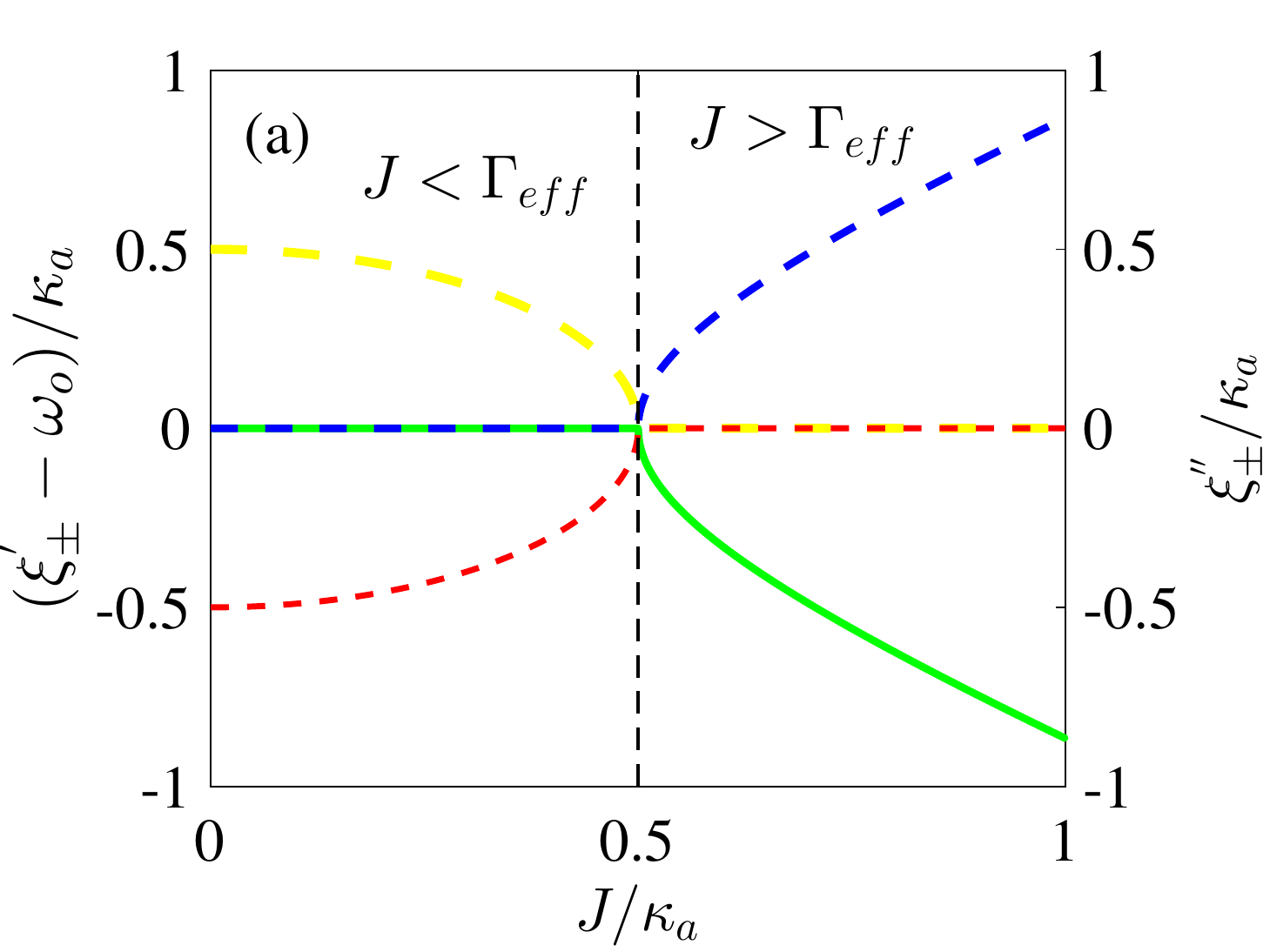}}
	\hspace{2mm}
	\subfigure{\includegraphics[width=0.48\linewidth]{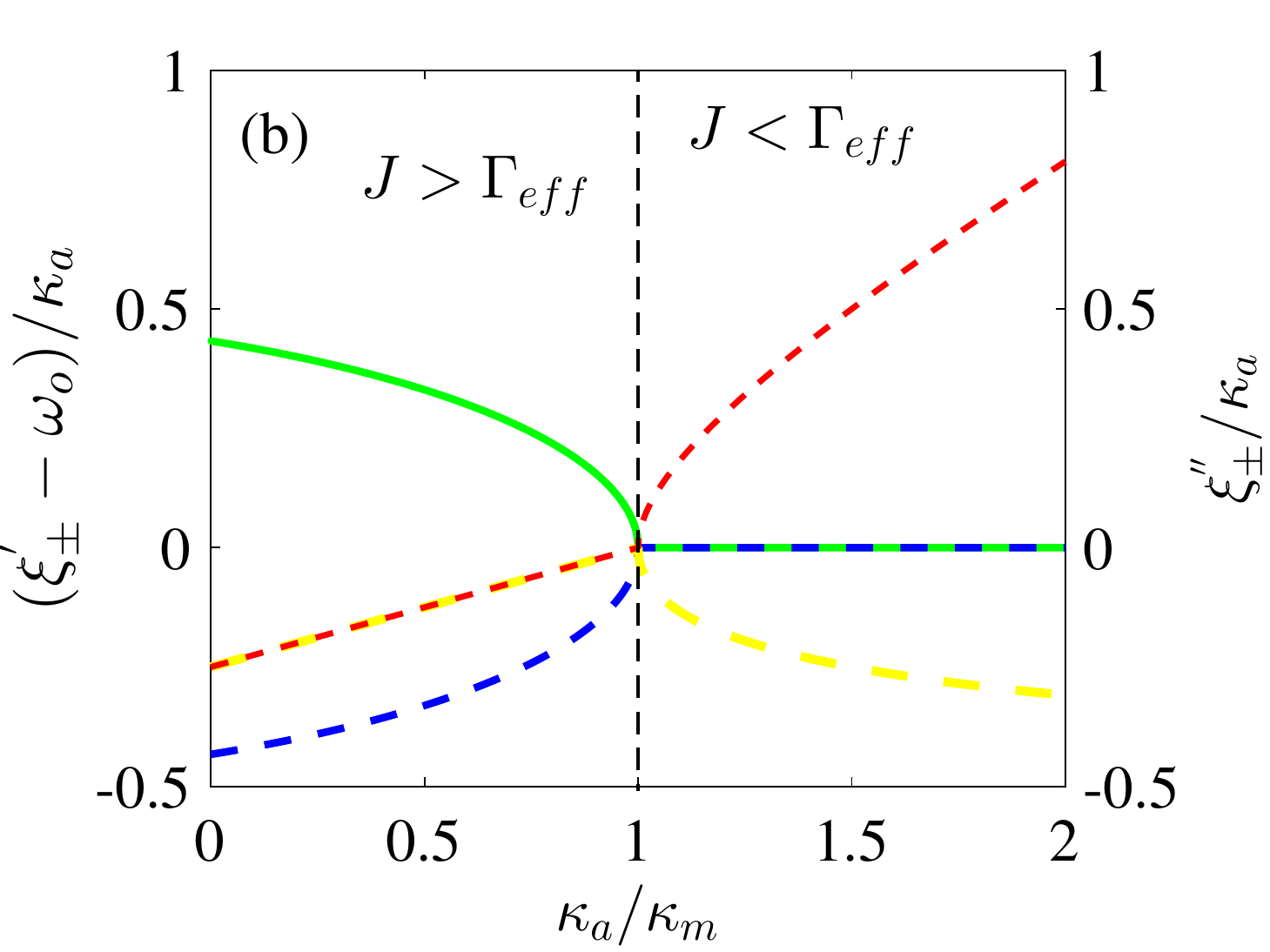}}
	\caption{(Color online) The real and imaginary parts of complex eigenfrequencies $\xi{\pm}$ (a) versus the magnon-photon coupling strength $J$ with balanced gain and decay rates of the system, i.e. $\kappa_{a}=\kappa_{m}$, and (b) versus the gain rate $\kappa_{a}$ with $J/\kappa_{m}=0.5$. The real and imaginary parts of complex eigenfrequencies $\xi{\pm}$ are denoted by $\xi^{'}_{\pm}$ (described by the green solid and blue dashed lines) and $\xi^{''}_{\pm}$ (described by the yellow and red dasned lines), respectively.}
	\label{fig2}
\end{figure}

In Fig.~\ref{fig2} (a), we plot the real ($\xi^{'}_{\pm}$) and imaginary ($\xi^{''}_{\pm}$) parts of the complex eigenfrequencies $\xi_{\pm}$ versus the magnon-photon coupling strength $J$ for the gain microwave cavity system. We can find that when the magnon-photon coupling strength $J$ is lesser than the effective loss $\Gamma_{eff}$ of the supermodes, i.e., $J<\Gamma_{eff}$ ($\kappa_{a}=\kappa_{m}$), the eigenfrequencies of two supermodes are degenerate, i.e., $\xi_{+}=\xi_{-}=\omega_{0}$, and the two supermodes have different linewidths $\chi$. On the contrary, when the magnon-photon coupling strength $J$ is larger than the effective loss $\Gamma_{eff}$, i.e., $J>\Gamma_{eff}$ ($\kappa_{a}=\kappa_{m}$), the frequencies of the two supermodes become nondegenerate but with the same linewidths, which approaches 0 when the gain and loss rates are balanced, i.e. $\kappa_{a}=\kappa_{m}$. In the case that the magnon-photon coupling strength is equal to the effective loss, i.e., $J=\Gamma_{eff}$ ($\kappa_{a}=\kappa_{m}$), which corresponds to the phase transition point, that is, the exceptional point (EP), the eigenfrequencies coalesce simultaneously and the linewidths become degenerate, meaning that the two degenerate supermodes possess the same linewidth. The regions where the magnon-photon coupling strength $J$ is larger or smaller than the critical effective loss value ($\Gamma_{eff}$) correspond to the unbroken and broken $\mathcal{PT}$-symmetry regions, respectively. The phase transition between the two regions can be achieved by changing the magnon-photon coupling strength $J$. In Fig.~\ref{fig2}(b), we plot the real ($\xi^{'}_{\pm}$) and imaginary ($\xi^{''}_{\pm}$) parts of the eigenfrequencies $\xi_{\pm}$ versus the gain rate $\kappa_{a}$ with $J/\kappa_{m}=0.5$. Obviously, by increasing the gain rate $\kappa_{a}$ in the gain microwave cavity, the dynamic characteristics of the two supermodes can be changed from the same decay rate of the two modes to one with loss and the other with gain. Also, we find that when the gain rate is larger than the loss rate, the real part of the eigenfrequencies is closed, while the imaginary part is separated.

\begin{figure}
	\centering
	\subfigure{\includegraphics[width=0.8\linewidth]{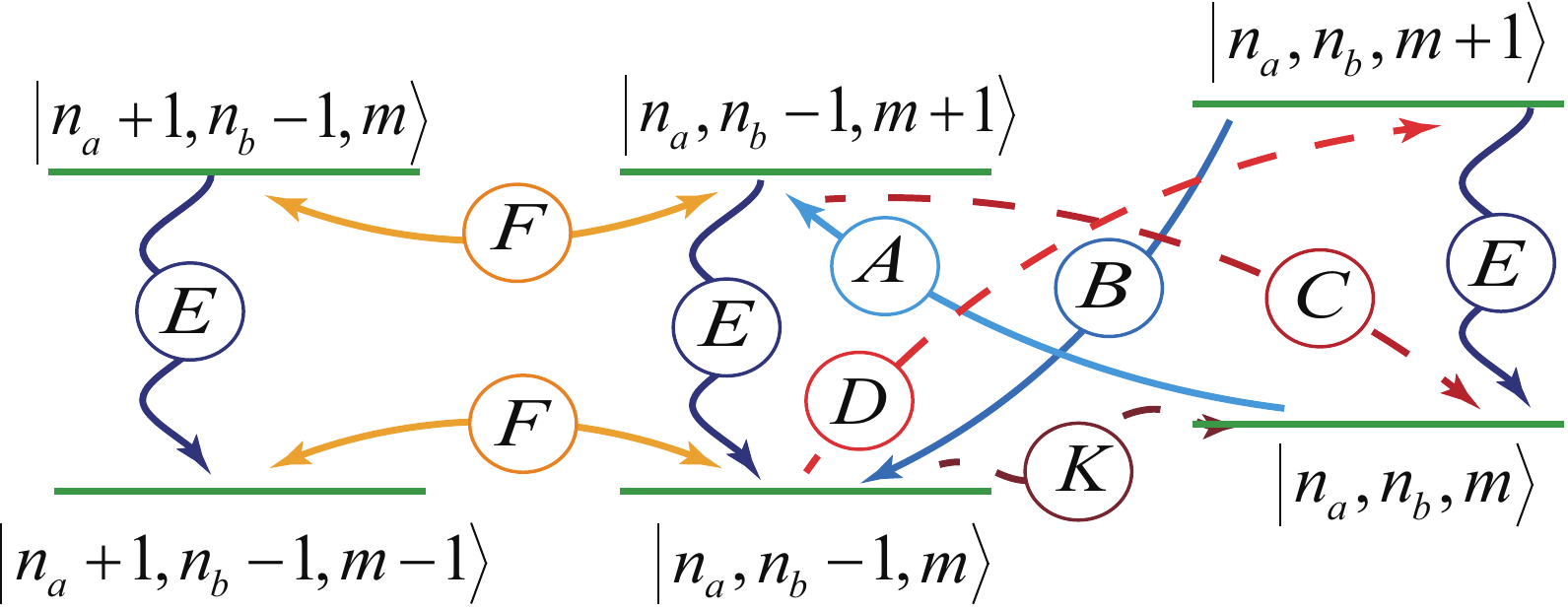}}
	\caption{ Energy level diagram of the linearized Hamiltonian in Eq.~(\ref{e03}), where $\arrowvert n_{a},n_{b},m\rangle$ denotes the number state, with $n_{a}$ being the photon number of microwave cavity mode, $n_{b}$ being the phonon number of magnomechanical resonator mode, and $m$ being the magnon number of magnon mode, respectively. Three blue solid lines $A$ (energy swapping), $B$ (counter-rotating-wave interaction), and $E$ (dissipation cooling) denote three different cooling processes. Three red dotted lines $C$ (swap heating), $D$ (quantum backaction heating), and $K$ (thermal heating) denote three different heating processes. The energy swapping due to magnetic-dipole interaction is denoted by the yellow solid  lines $F$.}
	\label{fig3}
\end{figure}

In Fig.~\ref{fig3}, we display the energy level diagram of the linearized Hamiltonian in Eq.~(\ref{e03}) and all the coupling routes in the displaced frame $| n_{a},n_{b}, m \rangle$, where $n_{a}$, $n_{b}$, and $m$ are the photon, phonon, and magnon numbers, respectively. Due to the magnomechanical interaction, different kinds of cooling and heating processes may occur. The cooling processes associated with the swap cooling, the counterrotating wave interaction between magnon and phonon modes, and the dissipation cooling via the magnon mode are described by the blue solid curves $A, B$, and $E$, respectively. The heating processes corresponding to the swap heating derived from the energy exchange between magnon and phonon modes, the quantum backaction heating that is the fundamental limit for backaction cooling, and the thermal heating that is an incoherent process are denoted by the red dotted curves $C, D$, and $K$, respectively. Note that the quantum backaction heating and swap heating are the accompanying effects, which correspond to the coherent interaction processes $\delta m^{\dagger}\delta b^{\dagger}$ and $\delta m\delta b^{\dagger}$, respectively. Here, the magnons backaction heating is different from the backaction heating of photons in the cavity  optomechanical systems. The photons quantum backaction heating is the accompanying effect when radiation pressure is utilized to cool the mechanical motion in the cavity optomechanical systems. The magnons quantum backaction heating is the accompanying effect when the magnetostrictive force is utilized to cool the magnon motion in the cavity  magnomechanical systems. Moreover, the energy swapping due to  magnetic-dipole interaction is represented by the yellow solid curve $F$. Our aim is to achieve efficient ground state cooling of magnomechanical resonator, which needs to enhance the cooling effect while suppressing the heating.

\section{\label{sec.3}Ground State Cooling Of Magnomechanical Resonator}
In this section, we discuss and study the ground state cooling of the magnomechanical resontor in the proposed cavity magnomechanical system. Similar to the method used in optomechanical systems, we first study the cooling rate of the magnomechanical resonator by using the quantum noise spectrum of the magetic force. It is well known that the force acting on magnomechanical resonator can be described by 
\begin{eqnarray}\label{e011}
F_{\mathrm{m}}\left(t\right)=m_{\mathrm{eff}} \ddot{x}+m_{\mathrm{eff}}\omega_{b}^{2}x,
\end{eqnarray}
and the equation of motion of the momentum operator $p$ is given by
\begin{eqnarray}\label{e012}
\dot{p}+m_{\mathrm{eff}}\omega_{b}^{2}x&=&-\frac{G^{*}\delta m\left( t\right) +G\delta m^{\dagger}\left( t\right) }{x_{\mathrm{ZPF}}},
\end{eqnarray} 
where $m_{\mathrm{eff}}$ is the effective mass of magnomechanical resonator, $x=x_{\mathrm{ZPF}}(\delta b^{\dagger}+\delta b)$ is the position operator, and $p=im_{\mathrm{eff}}\omega_{b}x_{\mathrm{ZPF}}(\delta b^{\dagger}-\delta b)$ is the momentum operator, with $x_{\mathrm{ZPF}} = \sqrt{\frac{1}{(2m_{\mathrm{eff}}\omega_{b})}}$ being the zero-point fluctuation amplitude of magnomechanical resonator. The commutation relation for operators $x$ and $p$ is $[x,p] = i$. From Eq.~(\ref{e012}), the magnetic force operator can be obtained as
\begin{eqnarray}\label{e013}
F_{\mathrm{m}}\left( t\right)  &=& -\frac{[G^{\ast}\delta m\left( t\right)  +G\delta m^{\dagger}\left( t\right) ]}{x_{\mathrm{ZPF}}}.
\end{eqnarray}
Applying the Fourier transform of the autocorrelation functions, the quantum noise spectrum of the magetic force is given by 
\begin{eqnarray}\label{e014}
S_{\mathrm{FF}}\left( \omega\right) &=&\int\left\langle F_{m}\left( t\right) F_{m}\left( 0\right) \right\rangle e^{i\omega \label{key}t}dt.
\end{eqnarray}

To obtain the expression of the magetic force noise spectrum $S_{\mathrm{FF}}(\omega)$, we transform Eq.~(\ref{e05}) to the frequency domain and have
\begin{eqnarray}\label{e015}
\frac{\delta a\left(\omega\right)}{\chi_{a}\left(\omega\right)}  &=&
-iJ\delta m\left(\omega\right) -\sqrt{\kappa_{a}}a_{\mathrm{in}}\left( \omega\right),
\cr\cr
\frac{\delta b\left(\omega\right)}{\chi_{b}\left(\omega\right)}  &=&
-i\left[G\delta m^{\dagger}\left(\omega\right)+G^{*}\delta m\left(\omega\right)\right] -\sqrt{\gamma_{b}}b_{\mathrm{in}}\left( \omega\right),
\cr\cr
\frac{\delta m\left(\omega\right)}{\chi_{m}\left(\omega\right) }  &=&
-iJ\delta a\left(\omega\right)-iG\left[\delta b^{\dagger}\left(\omega\right)+\delta b\left(\omega\right)\right]-\sqrt{\kappa_{m}}m_{\mathrm{in}}\left(\omega\right), 
\end{eqnarray}
where $\chi_{a}(\omega)$, $\chi_{b}(\omega)$, and $\chi_{m}(\omega)$ are the response functions of the mircrowave cavity mode, phonon mode, and magnon mode, respectively, with
\begin{eqnarray}\label{e016}
\chi_{a}\left(\omega\right)&=&\frac{1}{-i\left(\omega+\Delta_{a}\right)-\kappa_{a}/2},
\cr\cr
\chi_{b}\left(\omega\right)&=&\frac{1}{-i\left(\omega-\omega_{b}\right)+\gamma_{b}/2},
\cr\cr
\chi_{m}\left(\omega\right)&=&\frac{1}{-i\left(\omega+\Delta_{m}\right)+\kappa_{m}/2}. 
\end{eqnarray} 
Note that $\delta m(\omega)$ is the Fourier transform of $\delta m(t)$ and the important relation $\delta m^{\dagger}(\omega)=[\delta m(-\omega)]^{\dagger}$ should be taken into account while solving Eq.~(\ref{e016}). Then, by using $F_m(\omega)=-\frac{\hbar[G^{\ast}\delta m(\omega)+G\delta m^{\dagger}(\omega)]}{x_{\mathrm{ZPF}}}$ in the frequency domain, the quantum noise spectrum density of the magnetic force operator is calculated as
\begin{eqnarray}\label{e017}
S_{\mathrm{FF}}\left(\omega\right)  &=&
\frac{\left|G\right|^2}{x^2_{\mathrm{ZPF}}}[\gamma_{b}\left| \chi\left(\omega\right)\right|^2+\kappa_{a}J^2\left| \chi\left(-\omega\right)\right| ^2\left| \chi_{a}\left( -\omega\right)\right|^2],
\end{eqnarray}
where
\begin{eqnarray}\label{e018}
\chi\left( \omega\right) &=&\frac{\chi_{m}(\omega)}{1+J^2\chi_{a}(\omega)\chi_{m}(\omega)+|G|^2\chi_{b}(\omega)\chi_{m}(\omega)},
\end{eqnarray}
which represents the total response function of the cavity magnomechanical system.

\begin{figure}
\centering
\subfigure{\includegraphics[width=0.49\linewidth]{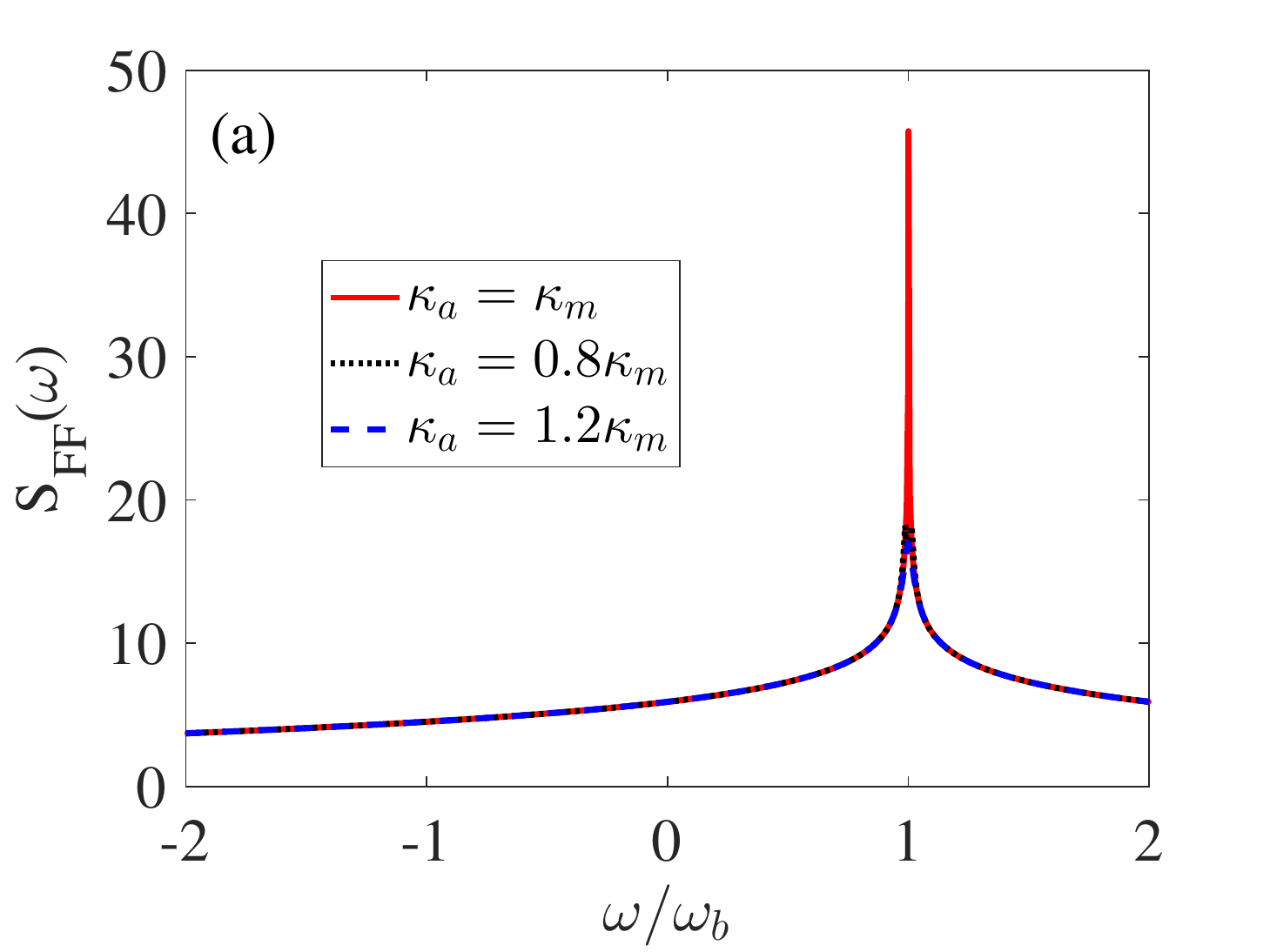}}
\hspace{-1mm}
\subfigure{\includegraphics[width=0.49\linewidth]{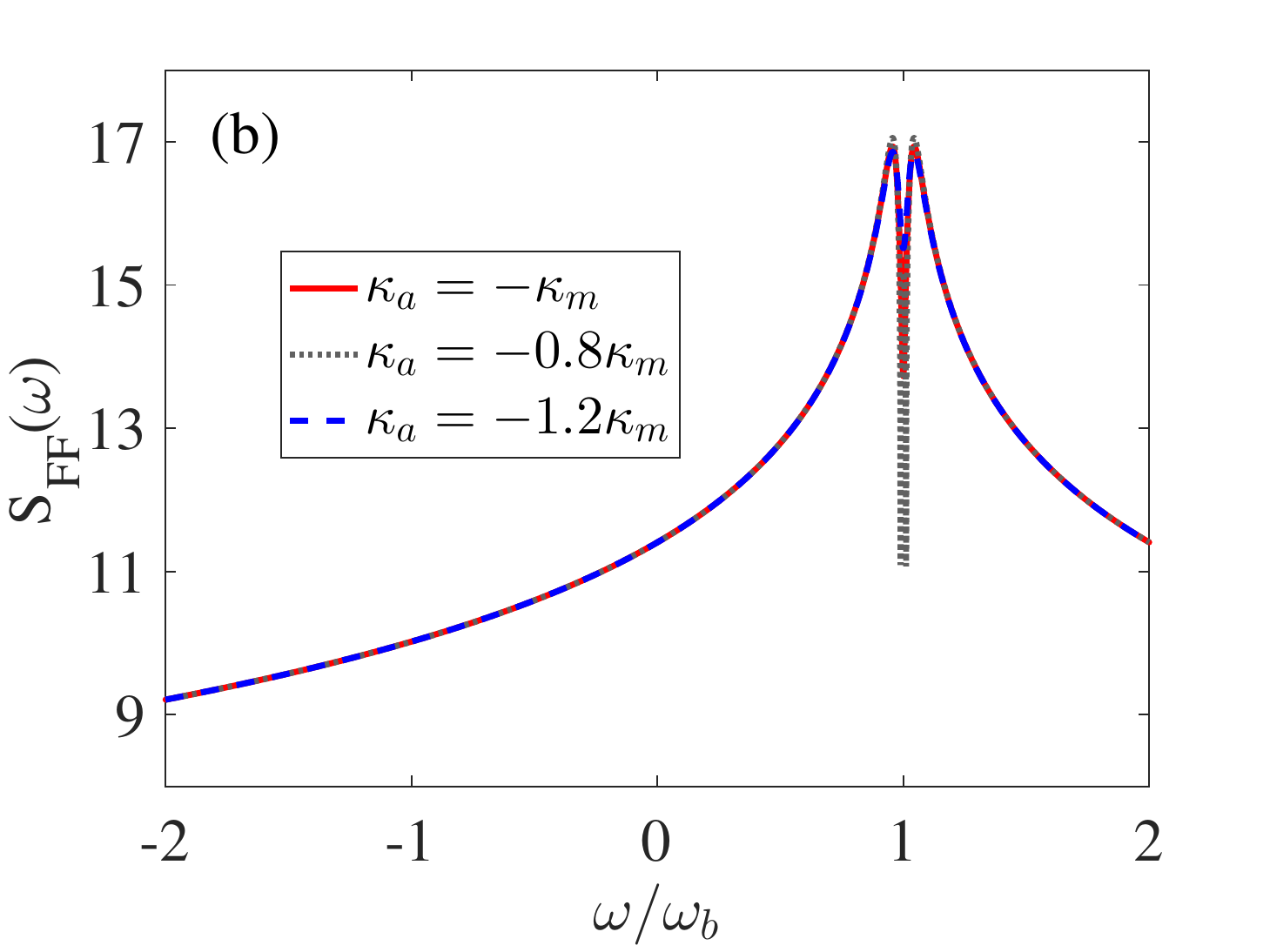}}
\caption{ (Color online) The magnetic force noise spectrum $S_{\mathrm{FF}}(\omega)$ as a function of the frequency $\omega$ for (a) the gain microwave cavity system and (b) the loss microwave cavity system. For the two different microwave cavity systems, three different gain (decay) rates of the cavity mode $\kappa_{a}=0.8\kappa_{m}~(-0.8\kappa_{m})$ (black dashed line), $\kappa_{a}=\kappa_{m}~(-\kappa_{m})$ (red solid line), and $\kappa_{a}=1.2\kappa_{m}~(-1.2\kappa_{m})$ (blue dotted line) are considered. The other parameters are $\omega_{a}/2\pi=\omega_{m}/2\pi=10.1\mathrm{GHz}$, $\omega_{b}/2\pi=10 \mathrm{MHz}$, $G/\omega_{b}=0.03$, $\gamma_{b}/\omega_{b}=10^{-5}$, $\kappa_{m}/\omega_{b}=0.2$, $J/\omega_{b}=0.1$. Here, we set $\Delta_{a}=\Delta_{m}=\bar{\Delta} =-\omega_{b}$.}
\label{fig4}
\end{figure}

According to the Fermi golden rule, the heating and cooling rates correspond to the emission and absorption of phonons, which are calculated and given by  $A_{+}=S_{\mathrm{FF}}(-\omega)x^2_{\mathrm{ZPF}}$ and $A_{-}=S_{\mathrm{FF}}(\omega)x^2_{\mathrm{ZPF}}$, respectively. In order to cool the magnomechanical resonator to its ground state, the heating rate $A_{+}$ should be much lower than the cooling rate $A_{-}$, i.e., $A_{+}\ll A_{-}$. In Fig.~\ref{fig4}, we plot the magnetic noise spectrum $S_{\mathrm{FF}}(\omega)$ versus the frequency $\omega$ with differet gain (decay) rates. The effect of different gain rates on the cooling rate of magnomechanical resonator is shown in Fig.~\ref{fig4}(a), one can see that the position of the maximum value of the magnetic force noise spectrum $S_{\mathrm{FF}}(\omega)$ of the gain microwave cavity magnomechanical system is located at $\omega=\omega_{b}$, which corresponds to the maximum cooling rate. Moreover, the maximum value of the the magnetic force noise spectrum $S_{\mathrm{FF}}(\omega)$ can be significantly enhanced when the the gain and decay rates are balanced, i.e., $\kappa_{a}=\kappa_{m}$. Here the maximum cooling rate point corresponds to the phase transition point of the $\mathcal{PT}$-symmetry cavity magnomechanical system. For comparison, with the same system parameters, we also plot the magnetic force noise spectrum $S_{\mathrm{FF}}(\omega)$ of the system consisting of the loss YIG sphere and the loss microwave cavity, as shown in Fig.~\ref{fig4}(b). We can see that at the point $\omega=\omega_{b}$, the magnetic force noise spectrum $S_{\mathrm{FF}}(\omega)$ splits into two peaks, which originate from the mode coupling between the decay microwave cavity mode and the decay magnon mode. The spectral splitting results in the optimal cooling frequency shift and also reduces the maximum cooling rate. Comparing the two situations, we find that under the same system parameters, the maximum value of the magnetic noise spectrum in the $\mathcal{PT}$-symmetry cavity magnomechanical system is significantly increased at the position of $ \omega=\omega_{b} $. Therefore, a significantly enhanced cooling rate $A_{-}$ is obtained in our proposed cavity magnomechanical system. 

In the weak coupling regime ($G\ll\kappa_{a}$), the expression of $\delta b(\omega)$ can be obtained by solving Eq.~(\ref{e015}) (see Appendix~\ref{Appendix}). It is known that the imaginary part of $\sum(\omega_{b})$ corresponds to the net cooling rate of the magnomechanical resonator induced by the magnomechanical coupling,
  \begin{eqnarray}\label{e019}
 \Gamma=-2\mathrm{Im}\left[ \sum(\omega_{b})\right]=A_{-}-A_{+} ,
\end{eqnarray}
and the real part of $\sum(\omega_{b})$ corresponds to the frequency shift $\delta b(\omega)$ of magnomechanical resonator caused by the magnomechanical coupling,
\begin{eqnarray}\label{e020}
 \delta\omega_{b}=\mathrm{Re}\left[ \sum(\omega_{b})\right] ,
 \end{eqnarray}
For the detailed expressions and $\sum(\omega_{b})$, please see Appendix~\ref{Appendix}. In Fig.~\ref{fig5}, the net cooling rate $\Gamma$ of magnomechanical resontor is plotted as a function of the detuning $\bar{\Delta}$ for different gain and decay rates corresponding to two coupling strengths, $G/\kappa_{m}=0.15$ and 0.05, respectively. As shown in Fig.~\ref{fig5}(a), the maximum net cooling rate $\Gamma$ is located at the point $\bar{\Delta}=-\omega_{b}$. In Fig.~\ref{fig5}(b), the net cooling rete $\Gamma$ spilts into two peaks at the point $\bar{\Delta}=-\omega_{b}$, and the maximum value of the net cooling rate $\Gamma$ is less than that in the gain microwave cavity system. Moreover, the net cooling rate $\Gamma$ of the magnomechanical resontor is enhanced by about four orders of magnitude in the gain microwave cavity system. Also, we find that under the weak coupling regime ($G\ll\kappa_{a}$), the larger the magnomechanical coupling strength is, the larger the net cooling rate is.
 
\begin{figure}
	\centering
	\subfigure{\includegraphics[width=0.49\linewidth]{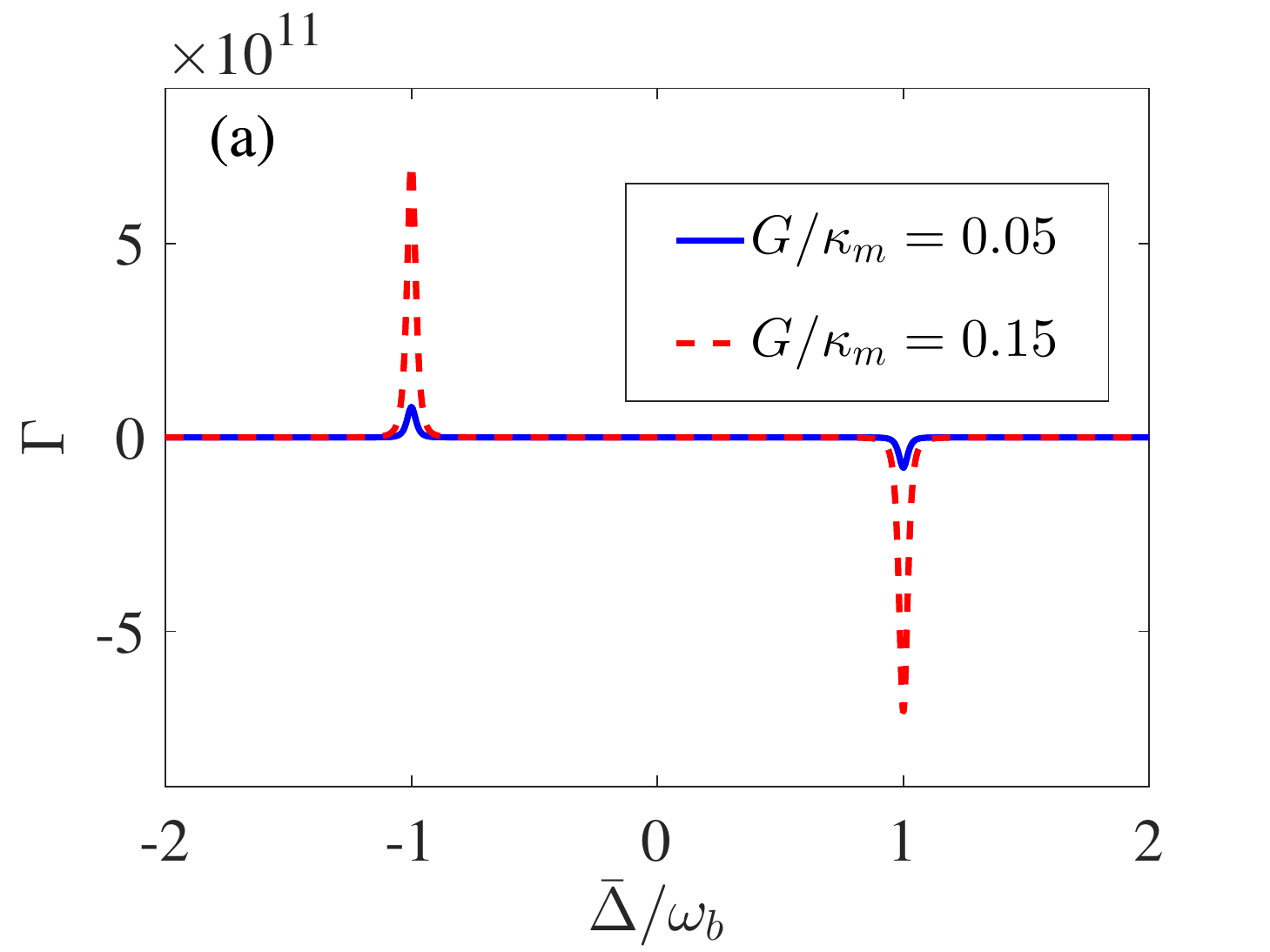}}
	\hspace{-1mm}
	\subfigure{\includegraphics[width=0.49\linewidth]{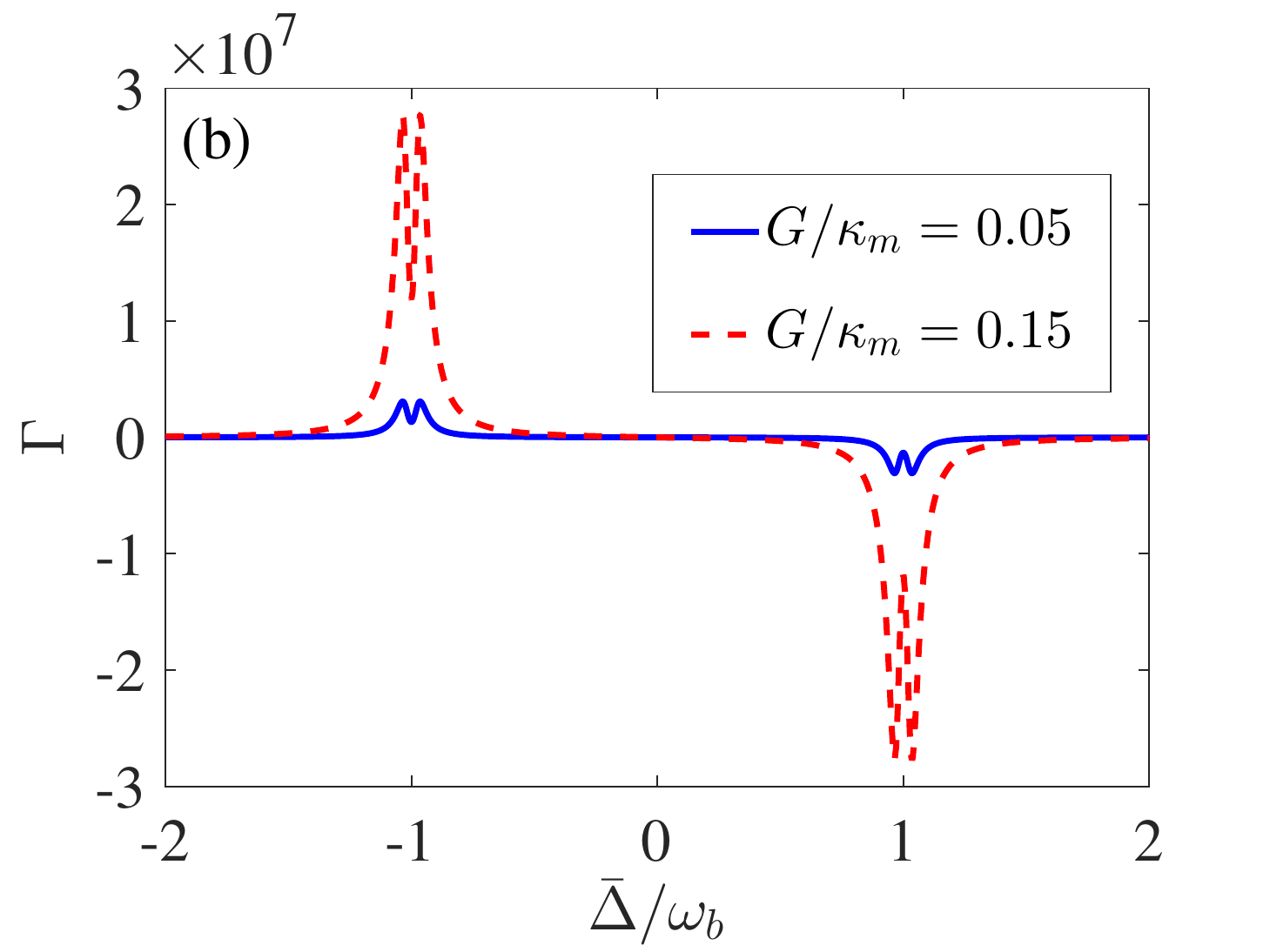}}
	\caption{(Color online) The net cooling rate $\Gamma$ verdus $\bar{\Delta}$ (a) the gain and the decay rates are balanced, i.e., $\kappa_{a}=\kappa_{m}$ and (b) the ratio of the two decay rates is a minus sign, i.e., $\kappa_{a}=-\kappa_{m}$ with two coupling strengths $G/\kappa_{m}=0.15$ (blue solid line) and $G/\kappa_{m}=0.05$ (red dashed line), respectively. The other parameters are chosen as the same in Fig.~\ref{fig4}. }
	\label{fig5}
\end{figure}
 
In the following, we investigate the final mean phonon number of magnomechanical resonator in the steady state. The rate equation for the probability $P_{n}(t)$ of the Fock state is written as  
\begin{eqnarray}\label{e021}
\dot{P_{n}}&=&
\left( A_{-}+\gamma_{b}\left( n_{\mathrm{th}}+1\right) \right) P_{n+1}+\left( A_{+}+\gamma_{b}n_{\mathrm{th}}\right) nP_{n-1}
\cr\cr&& 
-[ A_{-}n+A_{+}\left( n+1\right) +\gamma_{b}\left( n_{\mathrm{th}}+1\right) n\cr\cr&&+\gamma_{b}n_{\mathrm{th}}\left( n+1\right)] P_{n},
\end{eqnarray}
where $n$ is the phonon number of Fock state. In the steady state of the cavity magnomechanical system, the rate equation of the probability $P_n(t)$ can be solved by $\dot{\bar{n}} =0$, where $\bar{n}$ is the average phonon number denoted by $\bar{n}=\sum_{n=0}^{\infty}{nP_n}$. Thus the final mean phonon number $n_f$ is given by
\begin{eqnarray}\label{e022}
n_{f} &=&
\frac{\gamma_{b}n_{\mathrm{th}}+\Gamma n_{c}}{\gamma_{b}+\Gamma},
\end{eqnarray}
with
\begin{eqnarray}\label{e023} 
n_{c}&=&\frac{A_{+}}{A_{-}-A_{+}}.
\end{eqnarray}
Here $n_{c}$ is the quantum cooling limit, since when $\gamma_{b}$ $\rightarrow 0$, $n_{f}$ $\rightarrow$ $n_{c}$. Thus, $n_{f}$ can also be regarded as the steady state cooling limit, which can be divided into two parts: $(\textrm{i})$ the quantum cooling part $n^q_{f}=A_{+}/(\Gamma+\gamma_{b})$, corresponding to the heating rate $A_{+}$ that originates from the quantum backaction; $(\textrm{ii})$ the classical cooling limit $n^c_{f}=\gamma_{b}n_{\mathrm{th}}/(\Gamma+\gamma_{b})$. Therefore, the steady state cooling limit can be written $n_{f}=n^c_{f}+n^q_{f}$.

\begin{figure}
	\centering
	\subfigure{\includegraphics[width=0.49\linewidth]{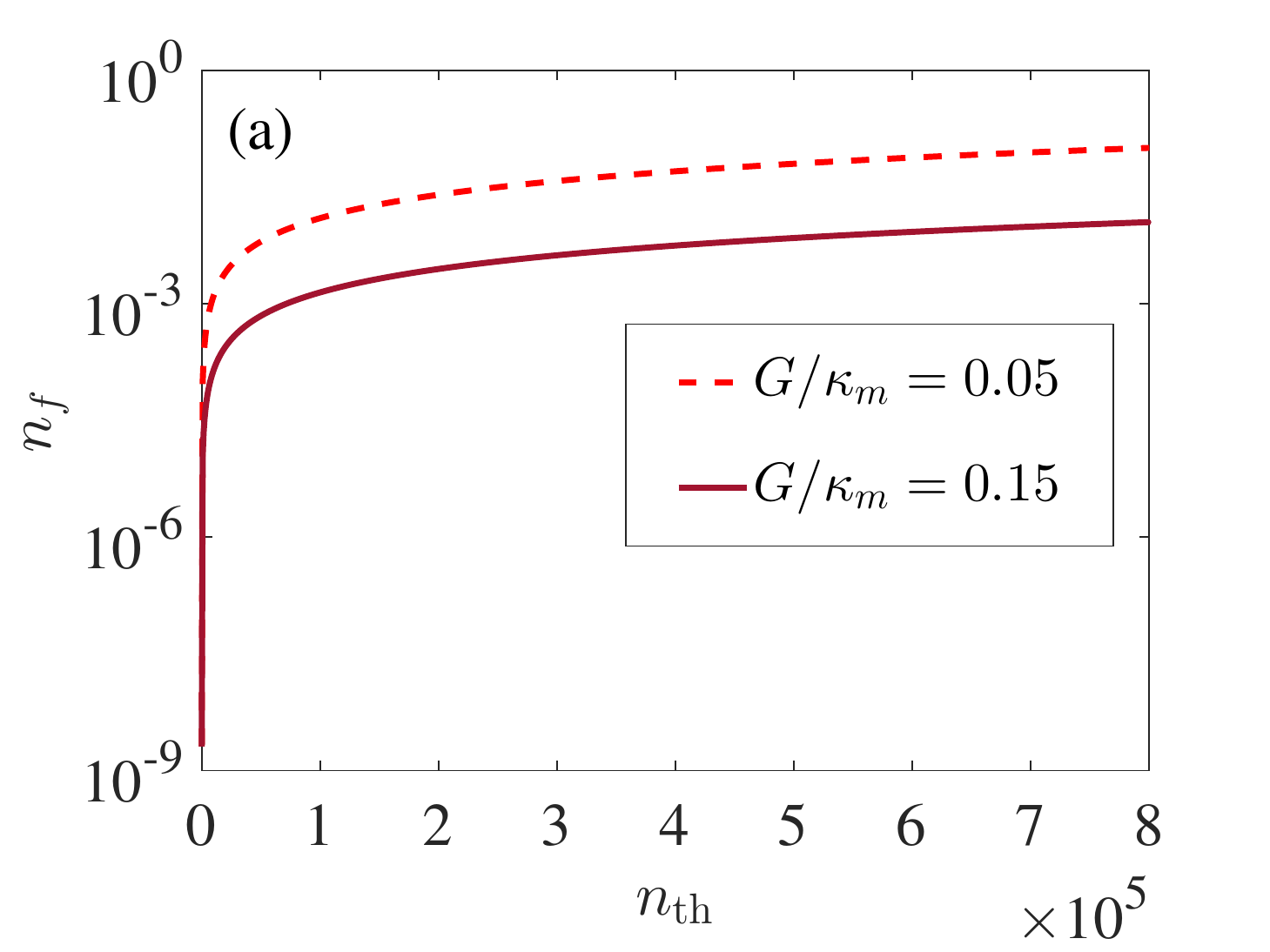}}
	\hspace{-1mm}
	\subfigure{\includegraphics[width=0.49\linewidth]{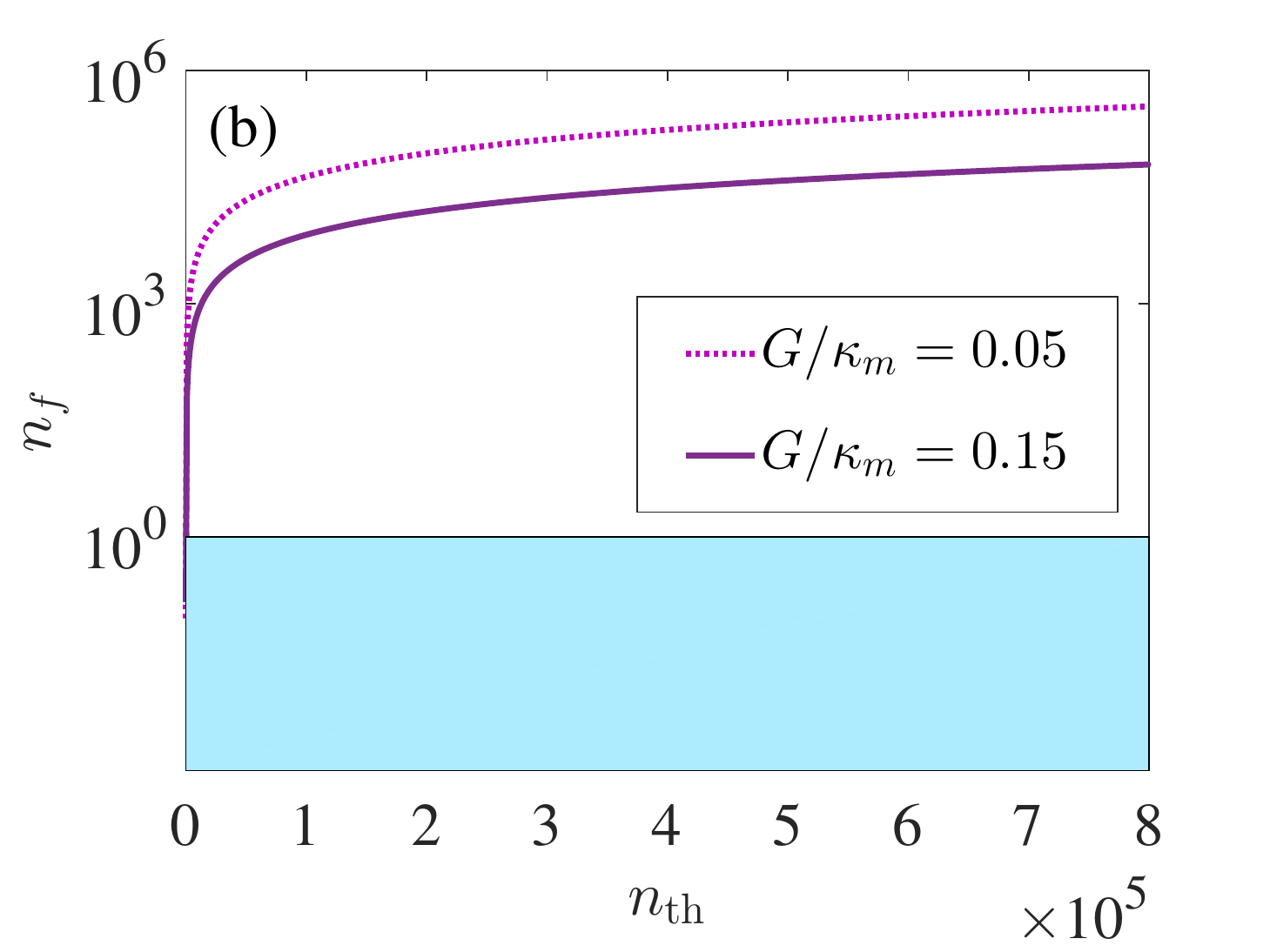}}
	\caption{(Color online) The final mean phonon number $n_{f}$ as a function of thermal phonon number $n_{\mathrm{th}}$ for (a) the gain microwave cavity system and (b) the loss microwave cavity system. The red solid line (purple solid line) and the light red dashed line (light purple dotted line) correspond to the coupling strengths $G/\kappa_{m} = 0.15$ and 0.05 for the gain magnomechanical system (the loss magnomechanical system), respectively. The other parameters are the same as in Fig.~\ref{fig4}.}
	\label{fig6}
\end{figure}

The cooling limit $n_{f}$ is mainly determined by the classical cooling limit $n^c_{f}$. From Eq.~(\ref{e019}) and Eq.~(\ref{e022}), we find that the final mean phonon number $n_f$ decreases with the increase of the net cooling rate $\Gamma$. For the gain microwave cavity system, the net cooling rate of magnomechanical resonator in Fig.~\ref{fig4}(a) is enhanced, which significantly reduces the classical cooling limit. In Fig.~\ref{fig6}(a), we show the final mean phonon number $n_{f}$ as a function of the thermal phonon number $n_{\mathrm{th}}$ with different coupling strengths $G/\kappa_{m}$ for the gain magnomechanical system. One can see that even if the initial thermal phonon number $n_{\mathrm{th}}$ is very large, the final mean phonon number $n_{f}$ is still less than 1.0. Interestingly, when the frequency of phonon mode is $\omega_{b}/2\pi=10~\mathrm{MHz}$, $n_{\mathrm{th}}=6.25\times10^5$, the corresponding temperature is $T=293~\mathrm{K}$, i.e., at the room temperature, which shows that the higher bath thermal phonon number  $n_{\mathrm{th}}$ can be tolerated for the ground state cooling of magnomechanical resonator in the gain microwave cavity, making that the initial cryogenic precooling of magnomechanical resonator can be greatly relaxed. Therefore, our work supports a precooling-free ground-state cooling of the magnomechanical resonator. For the loss cavity magnomechanical system, in Fig.~\ref{fig6}(b), in which the realization of ground state cooling is marked by the light blue shaded region (the final mean phonon number $n_{f}<1$). With the increase of the thermal phonon number $n_{\mathrm{th}}$, the final mean phonon number $n_{f}$ will soon exceed 1.0. Furthermore, to achieve the optimal final mean phonon number $n_{f,min}=1$ by the cooling in the loss cavity magnomechanical system, we choose the initial thermal phonon number as $n_{\mathrm{th}}=78$ ($G/\kappa_{m}=0.15$) and 830 ($G/\kappa_{m}=0.05$), in this case the corresponding bath temperatures are $T=40~\mathrm{mK}$ and $412~\mathrm{mK}$, respectively.

\begin{figure}
	\centering
	\subfigure{\includegraphics[width=0.49\linewidth]{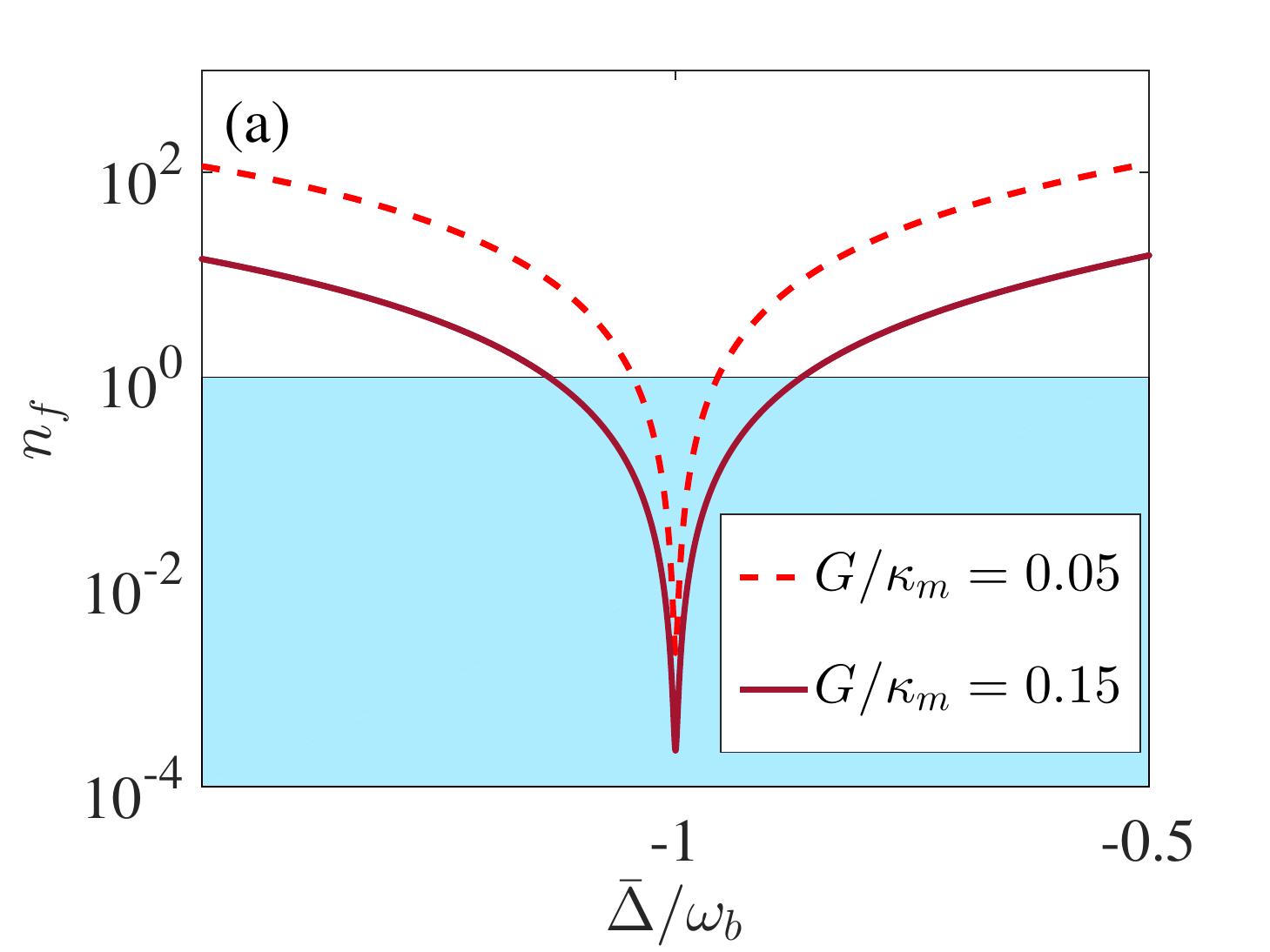}}
	\hspace{-1mm}
	\subfigure{\includegraphics[width=0.49\linewidth]{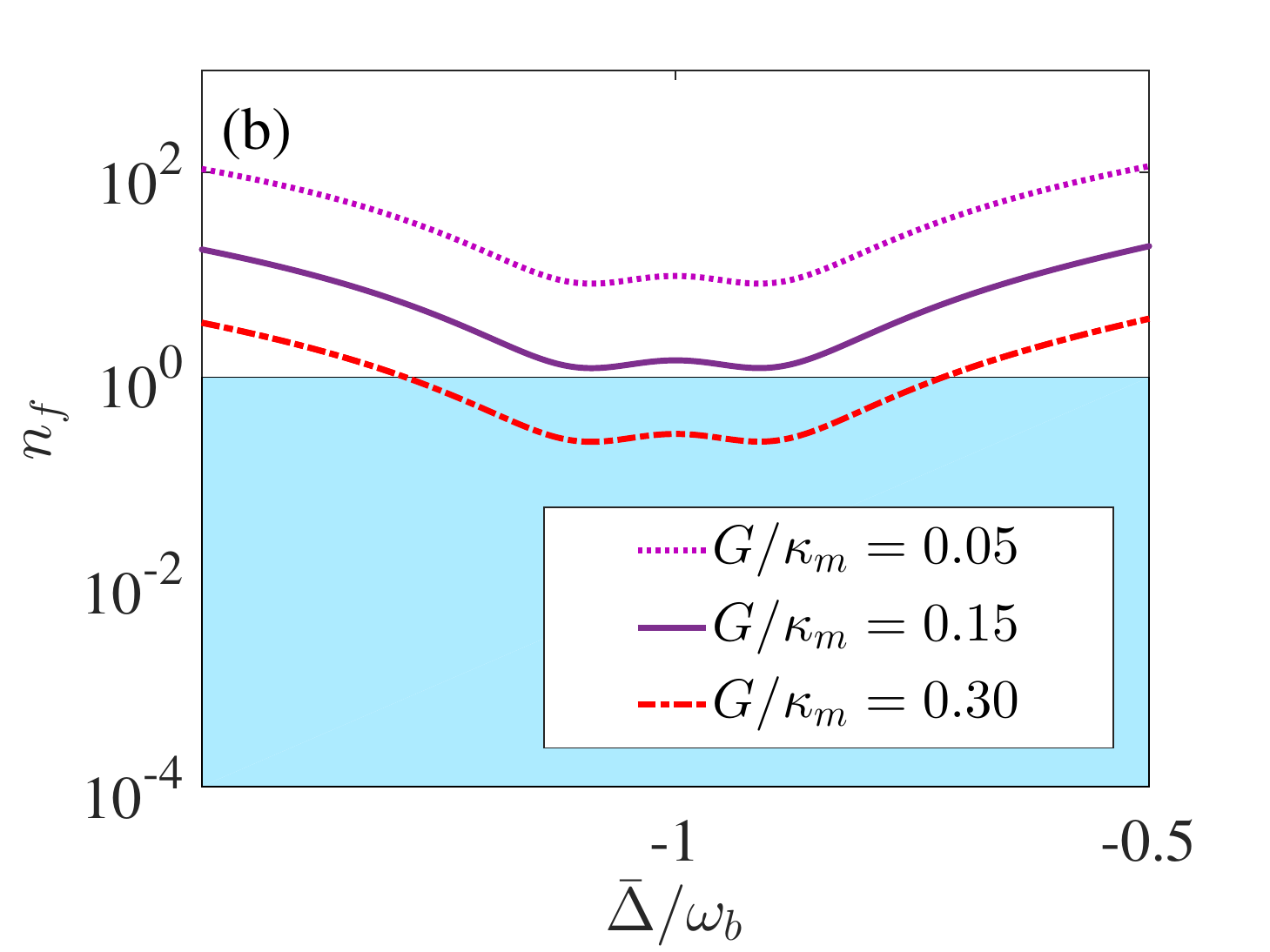}}
	\caption{(Color online) The final mean phonon number $n_{f}$ versus $\bar{\Delta}$ for (a) the gain microwave cavity system and (b) the loss microwave cavity system. The red solid line (purple solid line) and the light red dashed line (light purple dotted line) correspond to the coupling strengths $G/\kappa_{m} = 0.15$ and 0.05 for the gain magnomechanical system (the loss magnomechanical system), respectively. Here, the thermal phonon number is chosen as $n_{\mathrm{th}}=10^3$, and the other parameters are the same as in Fig.~\ref{fig4}.}
	\label{fig7}
\end{figure}

In Fig.~\ref{fig7}, we plot the final mean phonon number $n_{f}$ as a function of the cavity detuning $\bar{\Delta}$ with different coupling strengths $G/\kappa_{m}$ for the gain and loss cavity magnomechanical systems. We also mark the realization of ground-state cooling with light blue shaded region. One can see from Fig.~\ref{fig7}(a) that under the weak coupling regime ($G\leqslant\kappa_{a}$), the larger the coupling strength is, the better the ground state cooling effect is. And the minimum value of the final mean phonon number $n_{f}$ is located at $\bar{\Delta}=-\omega_{b}$ for the gain cavity magnomechanical system, $n_{f}\approx10^{-4}$ corresponds to $G/\kappa_{m}=0.15$, and $n_{f}\approx10^{-3}$ for $G/\kappa_{m}=0.05$, respectively. This shows that the ground state cooling of the magnomechaical resonator can be effectively achieved in the gain microwave cavity. In Fig.~\ref{fig7}(b), we find that the position of the optimal cooling at the point $\bar{\Delta}/\omega_{b}=-1$ shifts and splits into two dips for the loss cavity magnomechanical system, and the ground state cooling cannot be achieved under both the coupling strengths $G/\kappa_{m}=0.15$ and 0.05. Continuing to increase the coupling strength, e.g., $G/\kappa_{m}=0.30$, the ground state cooling of magnomechanical resonator can be achieved, which reveals that the magnomechanical coupling strength required for achieving the ground state cooling of magnomechanical resonator in the $\mathcal{PT}$-symmetrical situation is greatly reduced.

\begin{figure}
	\centering
	\subfigure{\includegraphics[width=0.5\linewidth]{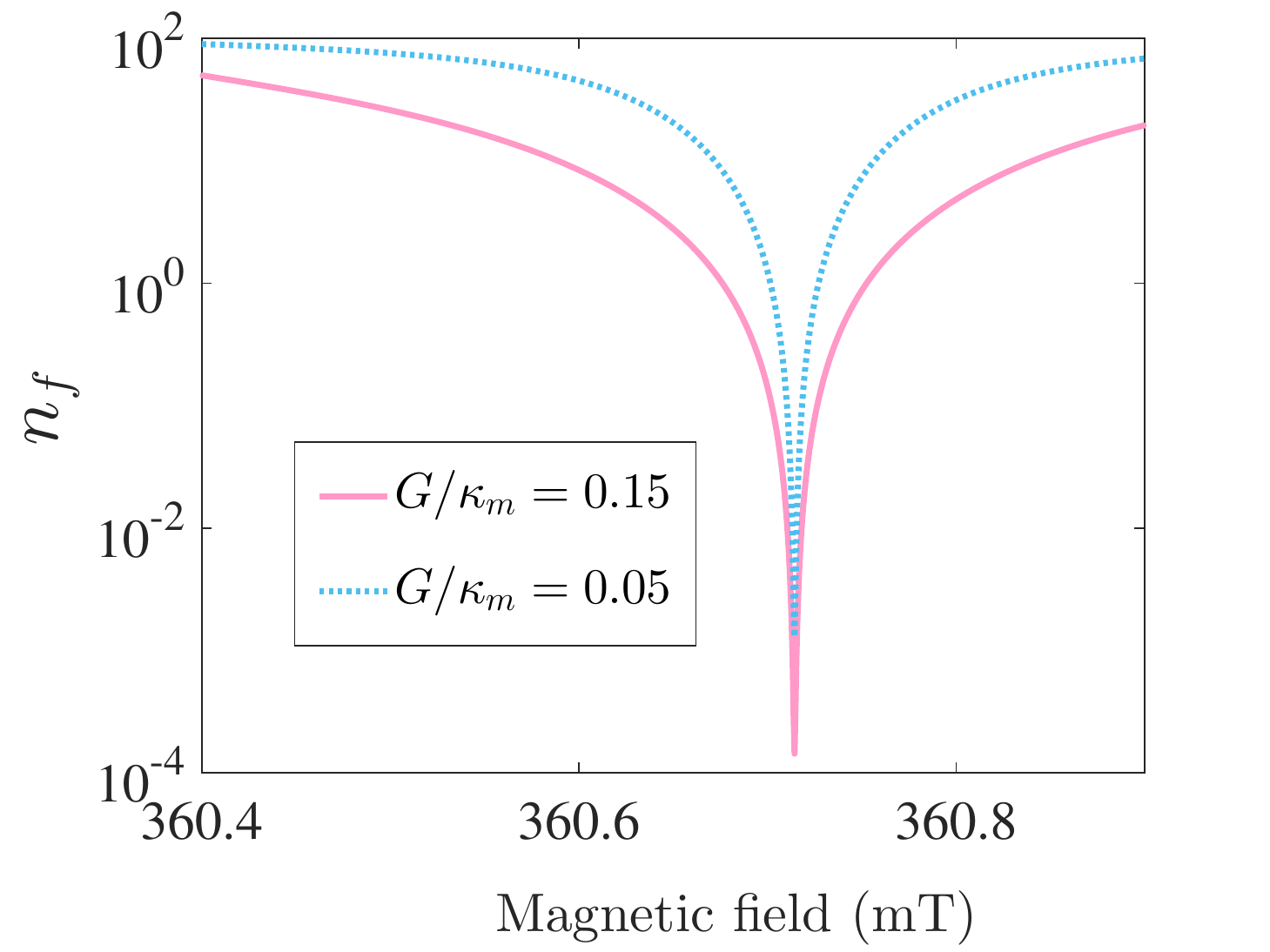}}
	\caption{The final mean phonon number $n_{f}$ versus the  magnetic field $H$ with different coupling strengths $G/\kappa_{m}=0.15$ (pink solid line) and $G/\kappa_{m}=0.05$ (blue dotted line) for the gain microwave cavity system. The other parameters are the same as in Fig.~\ref{fig3}.}
	\label{fig8}
\end{figure}

Now we investigate the influence of the magnon mode on the ground state cooling of magnomechanical resonator. The final mean phonon number $n_{f}$ versus the magnetic field $H$ with different coupling strengths is shown in Fig.~\ref{fig8}. We can see that when the magnetic field $H$ is close to 360.72~$\mathrm{mT}$, the final mean phonon number $n_{f}$ will approach the minimum value, which means that the ground state cooling is optimal. Moreover, the large magnomechanical coupling strength is beneficial to the realization of ground state cooling of the magnomechanical resonator. On the other hand, it is worth noting that the driving magnetic field, the bias magnetic field $H$, and the magnetic field of the cavity mode are mutually perpendicular at the site of the YIG sphere. We thus can control the ground state cooling of the magnomechanical resonator effectively by adjusting the magnetic field $H$ without changing other parameters. This property related to magnetic field may stimulate further exploration of the $\mathcal{PT}$-symmetrical cavity magnomechanical system.

\begin{figure}
		\centering
		\subfigure{\includegraphics[width=0.5\linewidth]{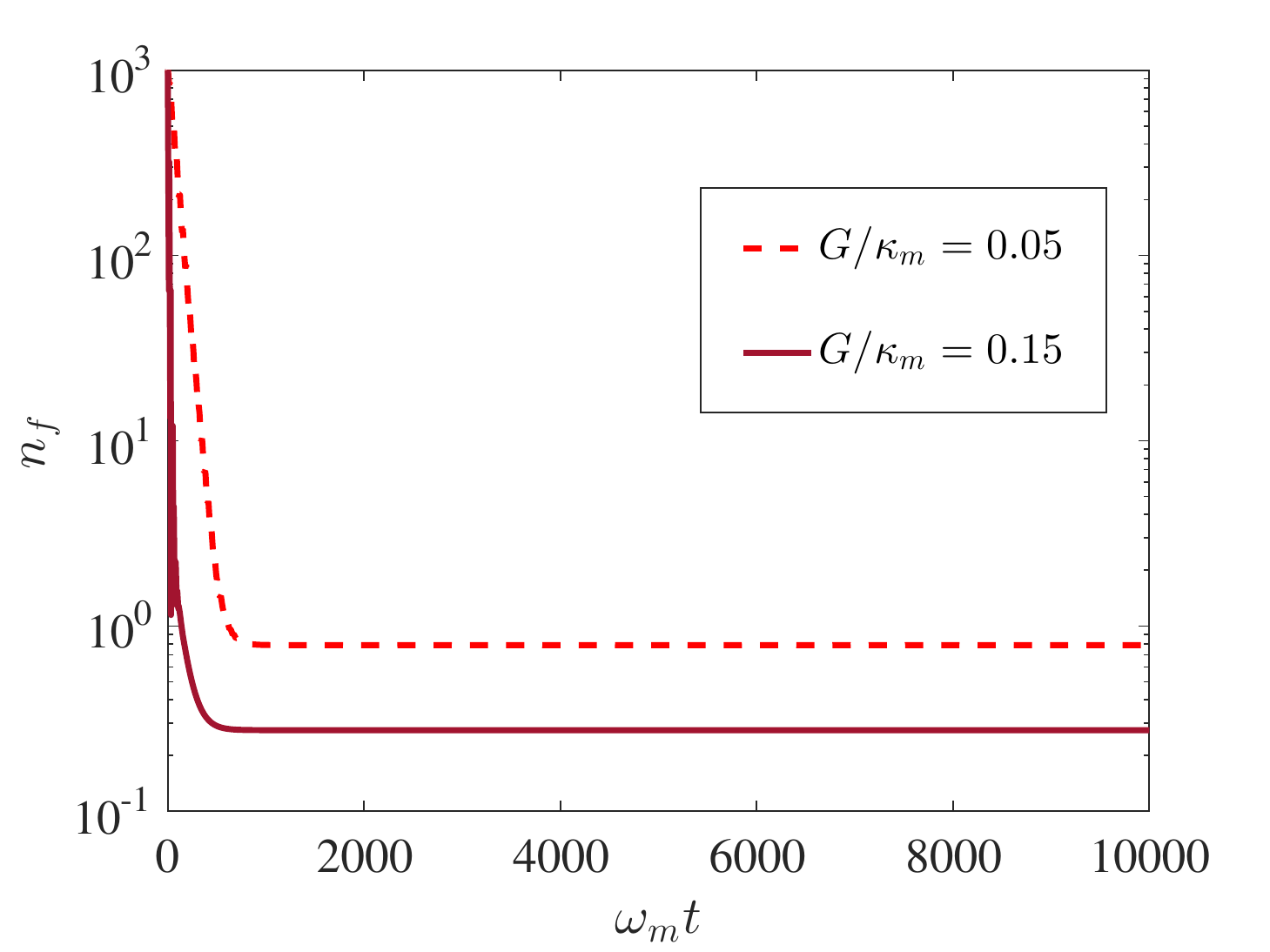}}
		\caption{(Color online) Time evolution of the final mean phonon number $n_{f}$ for the gain microwave cavity system. The red solid line and the light red dashed line correspond to the coupling strengths $G/\kappa_{m} = 0.15$ and 0.05 for the gain magnomechanical system, respectively. The other parameters are $\omega_{a}/2\pi=\omega_{m}/2\pi=10.1 \mathrm{GHz}$,  $\omega_{b}/2\pi=10 \mathrm{MHz}$, $G/\omega_{b}=0.03$, $\gamma_{b}/\omega_{b}=10^{-5}$, $\kappa_{m}/\omega_{b}=0.2$, $J/\omega_{b}=0.1$.}
		\label{fig9}
\end{figure}

We can find that for a given coupling strength $G$ of  magnon-phonon, there will always be a boundary of $G$ that divides the system into unstable region and stable region. Next, we will study the stability of the $\mathcal{PT}$-symmetric system. For the weak coupling regime (i.e., $G/\kappa_{m}=0.05$ and 0.15), we plot the numerical results of the final mean phonon $n_{f}$ based on covariance matrix for various $G$ in Fig.~\ref{fig9}. Here, we set the environmental phonon number as $10^3$. Initially, the phonon number of the magnon is equal to the environmental phonon number. As shown in Fig.~\ref{fig9}, we can see that with the increase of time, the final mean phonon occupancy in the magnon is cooled down to below one. It shows that for the weak coupling regime, the final mean phonon number decreases rapidly with the increase of the coupling strength.

\section{\label{sec.4}Conclusions}
In conclusion, we have studied a method to cool the magnomechanical resonator to its ground state in a $\mathcal{PT}$-symmetrical cavity magnomechanical system, which includes the microwave cavity mode, magonon mode, and phonon mode. The coupling between the magnon mode and the phonon mode is achieved through magnetostrictive interaction, and the magnon mode and the microwave cavity mode are coupled to each other via the magnon dipole interaction. We find that the $\mathcal{PT}$-symmetric phase transition can occur for the coupled cavity-magnon system in the gain microwave cavity by adjusting the magnon-photon coupling strength $J$. Compared with the non-$\mathcal{PT}$-symmetric system, the magnetic force noise spectrum in $\mathcal{PT}$-symmetric system is enhanced, the bandwidth is narrowed, and the net cooling rate around the phase transition point is increased. Moreover, the ground state cooling of the magnomechanical resonator in $\mathcal{PT}$-symmetric system does not need the cryogenic precooling, i.e., it is allowed to cool the magnomechanical resonator directly to its ground state at room temperature (293~$\mathrm{K}$). Furthermore, we can control the ground state cooling of the magnomechanical resonator by only adjusting the magnetic field without changing other system parameters. Our work provides a feasible method to cool the magnomechanical resonator and explore the quantum manipulation of macroscopic magnomechanical resonator at room temperature.

\begin{acknowledgments}
This work was supported by the National Natural Science Foundation of China under Grant No. 61822114.
\end{acknowledgments}

\appendix
\section{\label{Appendix} THE NET COOLING RATE}
By solving Eq.~(\ref{e07}), we can obtain the solution of $\delta b(\omega)$ as
\begin{eqnarray}\label{ae01}
\delta b\left( \omega\right)&=&\frac{\sqrt{\gamma_{b}}b_\mathrm{in}\left( \omega\right)-i\sqrt{\kappa_{m}}A\left( \omega\right) -\sqrt{\kappa_{a}}B\left( \omega\right)}{i\omega-i\left[ \omega_{b}+\sum\left( \omega\right) \right] -\gamma_{b}/2} ,
\end{eqnarray}
where 
\begin{eqnarray}\label{ae02}
A\left( \omega\right)  &=&G^{*}\chi\left( \omega\right) b_\mathrm{in}\left( \omega\right)+G\chi^{*}\left( -\omega\right) b^{\dagger}_\mathrm{in}\left( \omega\right),
\cr\cr
B\left( \omega\right)  &=&JG^{*}\chi\left( \omega\right)\chi_{a}\left( \omega\right)a_\mathrm{in}\left( \omega\right)\cr\cr&&-JG\chi^{*}\left(- \omega\right)\chi_{a}^{*}\left( -\omega\right)a^{\dagger}_\mathrm{in}\left( \omega\right),
\end{eqnarray}
and
\begin{eqnarray}\label{ae03}
\sum\left( \omega\right) &=&-i|G|^{2}\left[ \chi\left( \omega\right) -\chi^{*}\left( -\omega\right)\right],
\cr\cr
\chi \left( \omega\right) &=&\left[ J^{2}\chi_{a}\left( \omega\right) +\chi^{-1}_{m}\left( \omega\right)\right]^{-1},
\end{eqnarray}
here, $\sum(\omega)$ is the magnomechanical self energy and $\chi(\omega)$ is the total response function of the cavity magnomechanical system, and the term containing $b^{\dagger}(\omega)$ can be ignored because we consider $\omega\simeq\omega_{b}$. The magnomechaincal coupling and the net cooling rate are thus given by
\begin{eqnarray}\label{ae04}
\delta\omega_{b}&=&\mathrm{Re}\left[ \sum\left( \omega_{b}\right) \right] ,
\cr\cr
\Gamma&=&-2\mathrm{Im}\left[ \sum\left( \omega_{b}\right) \right] .
\end{eqnarray}



\end{document}